\documentclass[acus]{JAC2000US}

\usepackage{graphicx}

\begin{document}

%
\title{ A FAST SWITCHYARD FOR THE TESLA FEL-BEAM \\
USING A SUPERCONDUCTING TRANSVERSE MODE CAVITY }

\author{Rainer Wanzenberg, \\
DESY, Notkestr. 85, 22603 Hamburg, Germany }
\date{ }

\maketitle

\begin{abstract} 
In the present design of the TESLA Linear Collider with
integrated X-ray Laser Facility it is necessary that 1~ms long
bunch trains with about 10000 bunches are generated and
distributed to several free electron laser (FEL) beam
lines. The different scientific applications of the X-ray
FELs need specific filling patterns of the bunches in the
bunch train. It is shown that a fast switch-yard based on a
superconducting transverse mode cavity can be used to
generate the required bunch pattern in a flexible way
while keeping the beam loading in the main linear accelerator
constant. The conceptual design of the beam
optics and the transverse mode cavity are presented.
\end{abstract}


\section{Introduction}

\begin{figure}[h!btp]
\setlength{\unitlength}{1mm}
\centering
\includegraphics*[width=65mm]{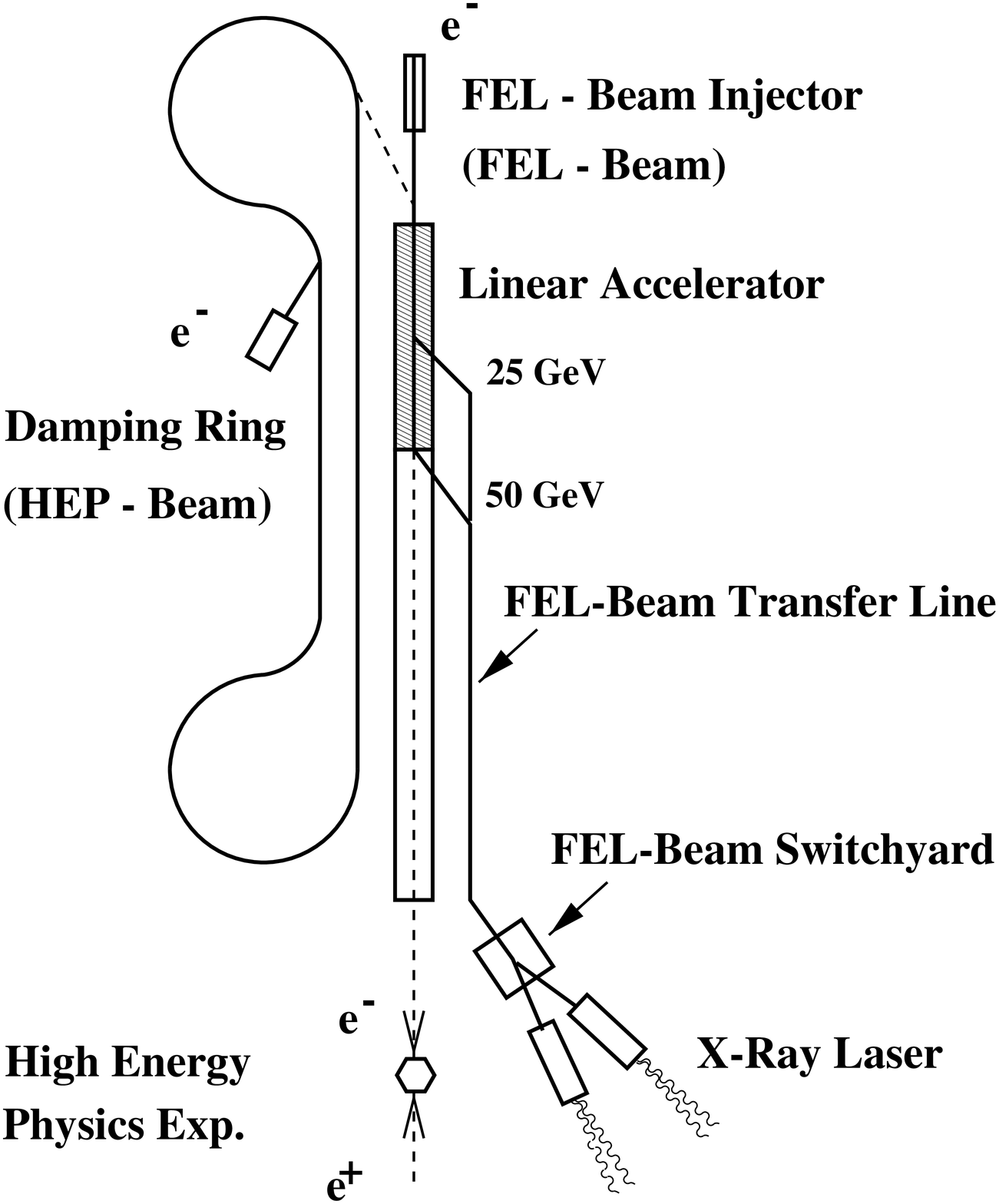}
\caption{The e$^{-}$ linear accelerator of the
TESLA linear collider with integrated x-ray laser
facility. }
\label{overview}
\end{figure}

\begin{figure}[h!btp]
\setlength{\unitlength}{1mm}
\centering
\includegraphics*[width=75mm]{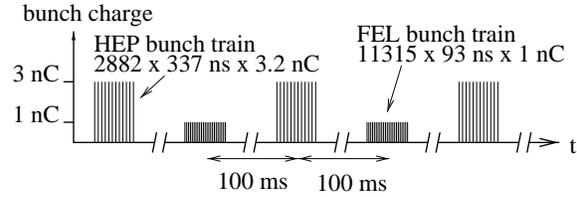}
\caption{HEP and FEL beam pulse structure}
\label{pulse}
\end{figure}

\begin{figure}[h!btp]
\setlength{\unitlength}{1mm}
\centering
\includegraphics*[width=60mm]{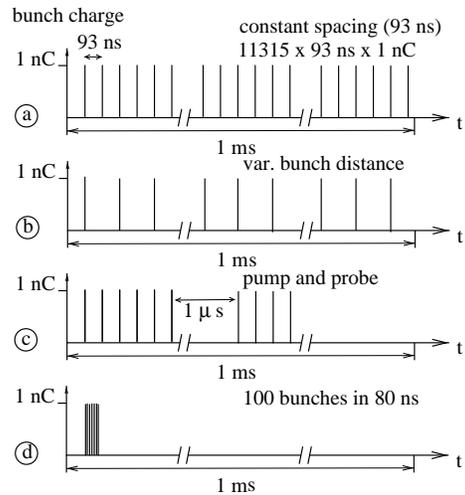}
\caption{Different bunch filling patterns of a FEL beam pulse.}
\label{felpulse}
\end{figure}

The conceptual design of the TESLA linear collider
with integrated x-ray laser facility \cite{CDR}
requires that 1~ms long bunch trains with 11315
bunches are generated and distributed to several
free electron laser (FEL) beam lines, while bunch trains
with 2882 bunches are accelerated to 250~GeV
for high energy physics (HEP) experiments.
The e$^{-}$ linear accelerator,
the two extraction points (at 25 GeV and 50 GeV) for the
FEL-beam and the beam transfer lines are shown
schematically in Fig.~\ref{overview}.
The first part of the e$^{-}$ linear accelerator is
operated at a duty cycle of 10~Hz providing
alternately  HEP and FEL  pulses.
 The pulse
structure is illustrated in Fig.~\ref{pulse}. The
mean pulse current is about 10 mA for the HEP and
FEL pulses, which guarantees the same beam loading
in the cavities for both pulse-types.
Different scientific applications of the X-ray FELs
need specific filling patterns of the bunches in the
FEL bunch trains \cite{Materlik}. Four examples of
filling patterns are shown in Fig.~\ref{felpulse}(a,b,c,d).
Case a is the (standard) 93~ns constant-spacing pattern,
while b and c are two examples how the number of bunches and the
bunch distance may be varied. Case d is a special case with
a much shorter bunch to bunch distance of 769 fs or
one $1.3$GHz rf-bucket. 
In the following sections it is shown how 
filling patterns like Fig.~\ref{felpulse}(b,c) 
can be generated from a standard constant-spacing pattern using
a fast switchyard based on a transverse mode cavity.
A much shorter bunch spacing as in Fig.~\ref{felpulse} (d)
of course requires a special bunch generation already at the
FEL beam injector. Whether such a bunch train can be accelerated
up-to 50 GeV without severe cumulative multi-bunch beam break-up
is beyond the scope of this paper.

\section{ Basic Design of a Fast Switchyard}

\begin{figure}[h!btp]
\setlength{\unitlength}{1mm}
\centering
\includegraphics*[width=75mm]{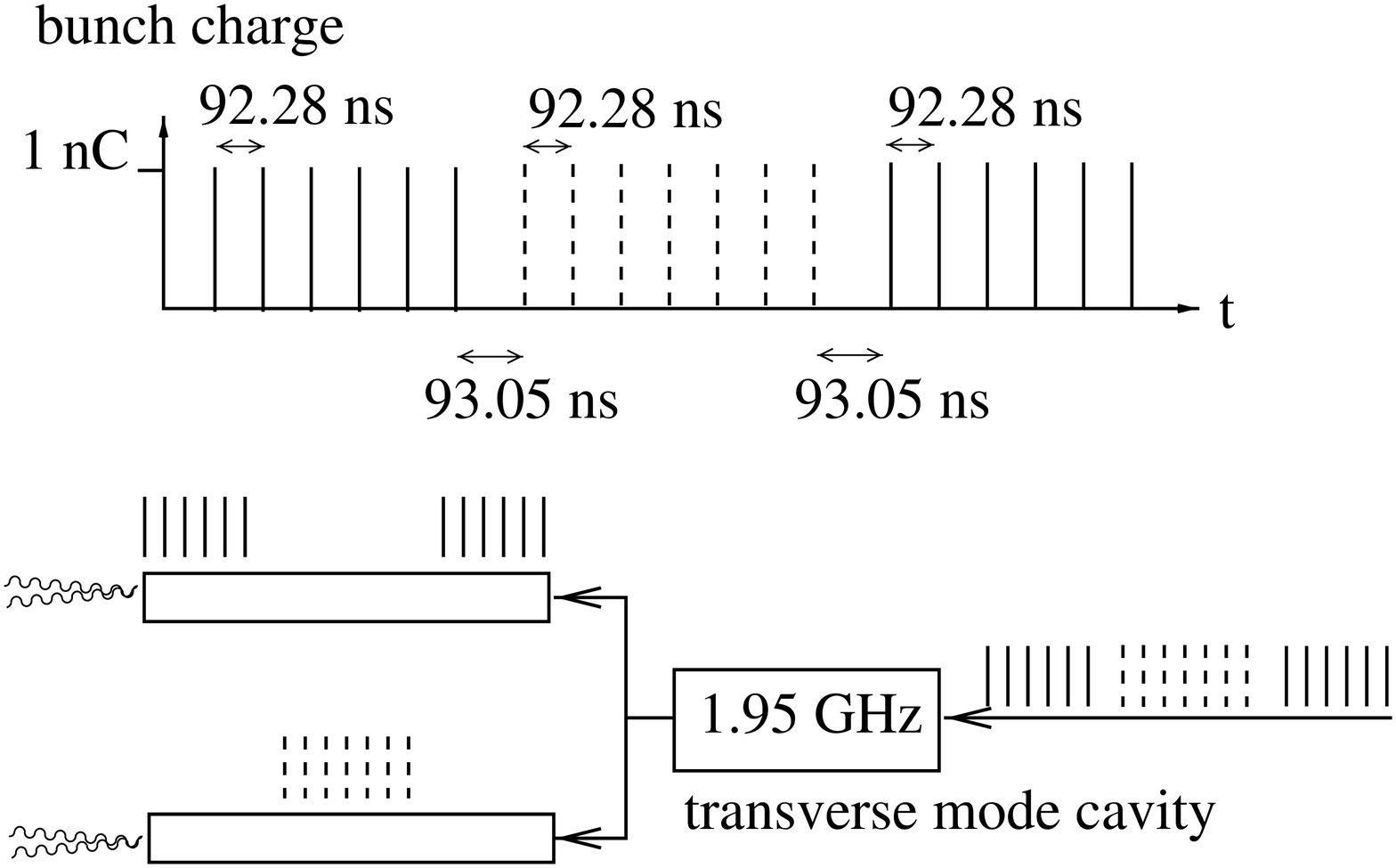}
\vspace*{-2mm}
\caption{Principle of a fast FEL-Beam switchyard}
\label{basics}
\end{figure}

The goal of a fast switchyard is to distribute single bunches 
or sub-trains of bunches within one 1~ms long bunch train to
different beam lines. The typical bunch distance is 120
rf-buckets of the $1.3$~GHz main linac rf-system or 
$120 \, \times \, 0.769 \, {\rm ns} = 92.28 \, {\rm ns}$. But
some scientific application of the FEL require special filling
patterns with even shorter and varying bunch distances (see Fig.~\ref{felpulse}).
This requirement can be accomplished by a pulsed superconducting
transverse mode cavity operated at a frequency of
$1.5 \times 1.3 \, {\rm GHz} = 1.95 \,  {\rm GHz}$ with a 1~ms rf-pulse
duration and a delay line for the laser system of the rf-gun.
The $1.95$ GHz deflecting cavity is operated in a pulsed mode
similar to the $1.3$~GHz accelerating cavity of the main linac.
This avoids rise time or stability problems of the kick applied to
individual bunches. The choice of the frequency labels the $1.3$~GHz
buckets as even and odd buckets. Bunches in even buckets are kicked into
the opposite direction than those in odd buckets, which enables the splitting of one
1~ms long pulse into several sub-bunch trains. The principle is 
illustrated in Fig.~\ref{basics}: A bunch train is generated with
a bunch-to-bunch distance of $92.28$~ns or 120 free $1.3$~GHz
buckets with a few exceptions where the distance is $93.05$~ns or
121 buckets. 
\vspace*{-2mm}
An even number of buckets between bunches guarantees that all bunches
are kicked into the same direction by the transverse mode cavity.
An odd number of free buckets between sub-bunch trains results in
a switch of the direction of the kick as show in Fig.~\ref{basics}.
The additional delay of one rf-bucket (or any odd number of rf-buckets)
can be achieved by an optical delay line of the laser beam pulse
at the rf-gun.

\begin{figure}[h!btp]
\setlength{\unitlength}{1mm}
\centering
\includegraphics*[width=75mm]{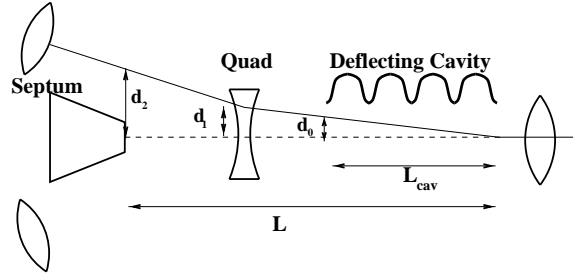}
\caption{Fast FEL-Beam Splitter. The kick due to
the defecting mode is enhanced by a defocusing
quadrupole magnet.}
\label{splitter}
\end{figure}

The beam optics of the switchyard is based on a FODO cell which is
shown in Fig.~\ref{splitter}. The kick due to the transverse mode
cavity is enhanced by a defocusing quadrupole \cite{Rossbach}.
A bunch offset $d_0$ of $ 5\, {\rm mm}$ at the end of the cavity
section,  $d_1 = 15\, {\rm mm}$ within the quadrupole and 
 $d_2 = 40\, {\rm mm}$ at the septum can be achieved with the
design parameters summarized in table~\ref{opticpara}
for two beam energies. In both cases a transverse gradient of
5 MV/m is necessary to provide a kick of $1.5$ ($1.0$) mrad.
The details of the cavity design are discussed in the next section.

A cascaded switchyard scheme with a $1.5 \times 1.3$~GHz and
additional $1.75 \times 1.3$~GHz transverse mode cavities would
allow the distribution of the bunches of a 1~ms long pulse
to four FEL beam lines. The details are not discussed in this
paper.
\vspace*{-5mm}
\begin{table}[htb]
\centerline{
\begin{tabular} {|l|c|c|}
        \hline
%
beam energy  & 25 GeV  &  50 GeV   \\
\hline
 total kick                       & $ 1.5 $ mrad & $1.0$ mrad \\
 active cavity length $L_{cav}$   & $ 7.5 $ m    & $10$  m    \\
 transverse gradient  $G_T$       & $ 5 $ MV/m   & $ 5 $ MV/m   \\
 total length (see Fig.~\ref{splitter}) $L$         & $ 17$ m      & $ 25 $ m \\
\hline
 quadrupole strength $k_{quad}$  & $0.2 \, {\rm m}^{-2}$ & $0.13 \, {\rm m}^{-2}$ \\
 min. beta function $\check{\beta}$ & $20$ m   & $28$ m \\
 max. beta function $\hat{\beta}$   & $61$ m   & $86$ m \\
\hline
\end{tabular}
}
\vspace{-1mm}
\caption{\label{opticpara} Design parameters of the
beam optics for the fast switchyard.}
\vspace*{-5mm}
\end{table}


\section{Design of the Transverse Mode Cavity}
\begin{figure*}[!htbp]
\setlength{\unitlength}{1mm}
\centering
\rotatebox{-90}{
\includegraphics*[width=22mm]{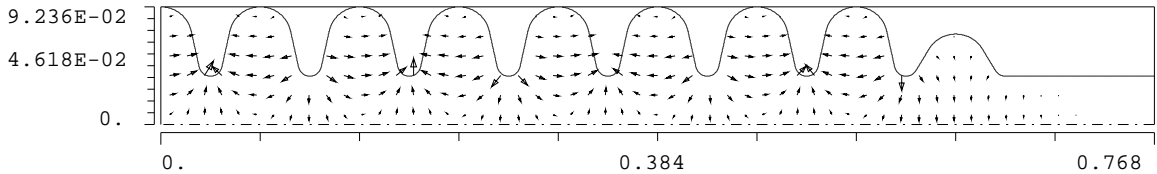}
}
\vspace*{-1mm}
\caption{Right half of the transverse mode cavity. 
The electric field of the $1.95$~GHz 
$\pi$-dipole-mode is shown (MAFIA calculation \protect\cite{MAFIA}).
15 cells contribute to the deflection, while the end-cells match the
field to that in the beam pipe.}
\label{pimode}
\vspace*{-4mm}
\end{figure*}

The basic design parameters of a transverse mode cavity
are the transverse gradient $G_T$, the peak magnetic field
on the surface $B_{peak}$, $(R/Q)'$ and $G_1$. The gradient
 $G_T$ is simply the average of the transverse component
of the Lorentz force $G_T = 1/L_{cav}
\int dz \, \left[ \, \vec{E_{\bot}}(z,t=z/c) +
               c \, \vec{e_z} \times \vec{B}(z,t=z/c)
           \right]$ acting on the beam; for a dipole mode
$G_T$ does not depend on the radial position of the beam
in the cavity.
Superconductivity breaks down when the rf magnetic field
exceeds the critical field of $0.2 \dots 0.24$~T for Niobium.
Therefore the transverse gradient $G_T$ is limited by
the peak magnetic surface field. 
A superconducting transverse mode S-band cavity has been operated
for an RF particle separator with a transverse gradient of
$1.2$ to $1.4$~MV/m \cite{KfK}. Present design studies of transverse
mode cavities at Fermilab \cite{Fermilab} are aiming at
gradients of $5.0$~MV/m.
An accelerating gradient
of 25~MV/m in the 1.3 GHz TESLA cavities corresponds to
a peak magnetic surface field of $0.105$~T. A similar peak magnetic
field of about $0.11$~T corresponds to a transverse gradient of
5~MV/m for the $\pi$-dipole-mode cavity shown in Fig.~\ref{pimode},
which represents one possible shape of a transverse mode cavity
with a relatively large iris diameter of 76~mm.
The results are obtained with the MAFIA \cite{MAFIA} code.
A large iris diameter is advantageous with respect to wakefield effects but
requires a special matching cell at the end of the cavity to
achieve good field flatness of the dipole mode.

\begin{table}[htb]
\centerline{
\begin{tabular} {|l|cc|}
%
\hline
 Frequency $f$           & $ 1.95 $ & GHz \\
 $(R/Q)'$                & $ 274  $ & Ohm \\
 $G_1$                   & $ 224  $ & Ohm \\
 Number of active cells  & $15$   &     \\
 Active length $L_{cav}$ & $1.15$ & m \\
\hline
 Transverse gradient $G_T$ & $5$ & MV/m \\
 Peak magnetic field $B_p$ & $0.11$ & T \\
\hline
 Q-value $Q_0$   & $3.8 \cdot 10^9$   & \phantom{\large{A}}   \\ 
  RF heat load (5~Hz, 2~K) &  $0.12$   & W  \\
\hline
 External Q-value $Q_{ext}$   & $3 \cdot 10^6$   & \phantom{\large{A}}   \\
 Filling time $T_F$      & $490$ & $\mu s$ \\
 RF-peak-power $P_{rf}$  & $20$    & kW \\
\hline
\end{tabular}
}
\vspace{-1mm}
\caption{\label{para} Basic design parameters of
the transverse mode cavity.  }
\end{table}

Further important parameters are $(R/Q)'$  and $G_1$, which
are defined according to the equations:
\begin{equation}
(R/Q)' = \frac{1}{4 \, \pi f} \,
         \frac{\left| G_T \, L_{cav} \right|^2}{U}, 
 \hspace*{3mm}
 Q_0 = \frac{G_1}{R_{BCS}(f,T)},
\end{equation}
where $U$ is the stored energy of the cavity mode and
$R_{BCS}(f,T)$ the BCS-resistivity of Niobium. The
parameter $(R/Q)'$ is essentially the ratio of the square
of the transverse gradient to the energy which is stored
in the cavity mode. $G_1$ is a purely geometrical parameter
which relates the surface resistivity to the Q-value of
the cavity. The BCS resistivity for the $1.95$~GHz cavity
at 2~K has been scaled from the $1.3$~GHz TESLA accelerating
cavity according to
\begin{equation}
 R_{BCS}(f,T) \sim (f^2 \,/T) \,
 \exp(-1.76 \, T_c/T),
\end{equation}
and using a Q-value of $1 \cdot 10^{10}$ for the TESLA
cavity. The dissipated power at 2~K during one pulse
for one transverse mode cavity
with an active length of $1.15$~m is 16~W according to
\vspace*{-5mm} 
\begin{equation}
P = \left( 5 \frac{ {\rm MV} }{ {\rm m}} \right)^2 \,
\frac{1}{2 \, (R/Q)' \, \, \, Q} \,\, ,
\end{equation}
with $Q_0 = 3.8 \cdot 10^9$, resulting in a average rf heat load
of $0.12$~W for a 5~Hz operation.
The same formula can be used
to calculate the required rf-peak-power by using the
external $Q$, $Q_{ext}$, which is determined by the coupling.
An external $Q$ of $3 \cdot 10^6$ has been chosen, for which
one obtains a filling time of $490 \,\, \mu s$
($T_F = Q_{ext}/(\pi \, f)$) which is similar to 
the filling time of the $1.3$~GHz TESLA accelerating cavity.

The switchyard for a 25 GeV (50 GeV) beam would require seven (ten) 
transverse mode cavities with the parameters considered in
table \ref{para}. The total rf-peak-power for 17 cavities is  
$340$~kW and the total rf heat load is 2~W at 2~K for a 5~Hz operation. 

\section{Conclusion}
It is feasible to distribute single bunches or
sub-bunch trains out of a 1~ms long bunch train to two
beam lines using a fast switchyard based on a transverse
mode cavity operated at $1.95$~GHz. The conceptual design
of the beam optics and the dipole mode cavity have been presented.
An engineering design of the system would require further
studies for the following subsystems: delay line of the laser
pulse at the rf-gun, integration of a dispersion suppression
and a collimation section into the beam optics, and design of
fundamental mode dampers at the transverse mode cavity. 
Depending on the required bunch pattern it is possible to
double the beam time for scientific applications
(e.g. pump and probe experiments) with a fast switchyard.
\begin{center}
{\bf Acknowledgments }
\end{center}
\vspace*{-1mm}
{\small
I would like to thank J.~Rossbach for discussions and
contributing ideas for the basic design of a fast
switchyard. Thanks go also to H.~Edwards and M.~McAshan
for their kind hospitality during my visit at Fermilab in
1999 where I became involved in the design of transverse mode
cavities. Furthermore I would like to thank
M.~Lomperski for carefully reading the manuscript during
breakfast.}


\end{document}